\documentclass[aps,preprint,showpacs]{revtex4}
\usepackage{graphicx,epsfig}
\usepackage{amssymb}
\usepackage{xcolor}
\begin{document}

\title{Red/Blue shifts of photons in a Hayward black hole immersed in perfect fluid dark matter}

\author{L.~Marlene.~L\'opez}
\email{lo281479@uaeh.edu.mx}
\author{L.~A.~L\'opez}
\email{lalopez@uaeh.edu.mx (Corresponding-Author)}

\affiliation{ \'Area Acad\'emica de Matem\'aticas y F\'isica, UAEH, 
Carretera Pachuca-Tulancingo Km. 4.5, C P. ~42184, Mineral de la Reforma, Hidalgo, M\'exico.}

\begin{abstract}
We investigate the red/blue shifts of photons emitted by particles moving along stable circular orbits around a Hayward black hole immersed in perfect fluid dark matter (PFDM). The PFDM contribution modifies the horizon structure and the characteristic radii associated with circular motion, such as the photon-sphere radius and the innermost stable circular orbit (ISCO). While massive particles follow geodesics of the background spacetime, photon propagation is described by an effective optical metric arising from nonlinear electrodynamics (NLED), which we use to determine the photon sphere and the critical impact parameter. The corresponding redshift and blueshift are sensitive to both the magnitude and sign of the PFDM parameter, providing potentially observable signatures of the surrounding dark matter distribution. Finally, we analyze the black hole shadow and compare its size with observational constraints from Sgr A*. Our results show that positive values of the PFDM parameter decrease the shadow radius, whereas negative values generally lead to larger shadows. The comparison with the observational bounds indicates that only certain negative ranges of the PFDM parameter are consistent with the data, suggesting that black hole shadow measurements can constrain the surrounding dark matter distribution.
\end{abstract}

\pacs{04.70.-s, 11.27.+d,98.62.Gq}
\maketitle

\section{Introduction}

Motivated by the limiting-curvature hypothesis, Hayward \cite{Hayward:2005gi} constructed a static and spherically symmetric regular black hole (BH) in which the central singularity is replaced by an effective de Sitter core, so that the curvature remains finite at the center. The Hayward BH can also be obtained from a generic class of solutions of general relativity coupled to nonlinear electrodynamics \cite{PhysRevD.94.124027}. Such regular solutions have been proposed as alternatives to classical black holes in order to address the singularity problem in general relativity.

Various extensions of the Hayward solution have been developed to explore a wide range of space-time properties. These include studies of the quasinormal modes of the Hayward black hole \cite{Lopez:2018aec,Lin:2013ofa}, analyzes that incorporate rotation \cite{Amir:2015pja}, and investigations of the Hayward black hole surrounded by quintessence \cite{Pedraza:2020uuy,Al-Badawi:2023lke,Yashwanth:2024suw}, among others.

The quasinormal modes of the Hayward black hole immersed in perfect fluid dark matter (Hayward--PFDM) have recently been investigated \cite{Tovar:2025apz}. This class of models is particularly relevant because it allows one to explore how the presence of dark matter -- estimated to constitute about $27\%$ of the total energy content of the Universe --modifies observable black hole properties.

The red/blue shifts of photons emitted by particles orbiting compact objects provide an effective tool for probing black hole properties and for determining physical parameters from directly observable quantities. Previous studies have shown that these frequency shifts can be used to extract information about the black hole mass, spin \cite{Herrera-Aguilar:2015kea}, and charge \cite{Becerril:2025wux}. The redshift and blueshift of a black hole immersed in a strong magnetic field \cite{Lopez:2021nud}, as well as those of black holes in nonlinear electrodynamics \cite{Lopez:2020ovf,dePaula:2025emo}, have also been studied. In a related context, a charged black hole immersed in PFDM has been analyzed to study the red/blue shifts of light emitted by particles orbiting the black hole \cite{Trad:2025xyi}.

In addition to frequency shifts, black hole shadows have become one of the most important observational tools for testing gravity in the strong-field regime. The Event Horizon Telescope (EHT) observations of M87$^{*}$ (2019) \cite{EventHorizonTelescope:2019dse} and Sgr A$^{*}$ \cite{EventHorizonTelescope:2022wkp} provide direct information about the space-time near black holes. Since the shadow boundary is determined by unstable photon orbits, it is highly sensitive to modifications of the gravitational field produced by surrounding matter distributions.

Motivated by the above considerations, we analyze the red/blue shifts of photons emitted by particles moving along stable circular orbits around a Hayward--PFDM black hole. In particular, we investigate how the PFDM parameter modifies the horizon structure, the characteristic radii of circular motion, and the resulting observable frequency shifts. We also investigate the shadow of the Hayward--PFDM black hole.

This article is organized as follows. In Sec.~II, we introduce the Hayward--PFDM black hole solution, briefly analyze its horizon structure, and discuss nonlinear electrodynamics, including the effective metric governing photon propagation. In Sec.~III, we examine the effective potential for massive test particles, while in Sec.~IV, we analyze the characteristic circular orbits. In Sec.~V, we determine the frequency shifts of photons emitted by particles moving along stable circular geodesics. In Sec.~VI, we investigate the shadow of the Hayward--PFDM black hole and discuss its dependence on the model parameters. Finally, the conclusions are presented in Sec.~VII.

\section{Hayward black hole immersed in perfect fluid dark matter}

Using the limiting curvature hypothesis and a minimal regularization scheme, Hayward~\cite{Hayward:2005gi} introduced a regular, static, spherically symmetric space-time intended to represent a regular black hole. In this construction, the Einstein tensor approaches an effective cosmological term in the central region, $G_{\mu\nu} \sim -\Lambda g_{\mu\nu}$ as $r \to 0$, with $\Lambda = 3/\epsilon^{2}$. The parameter $\epsilon$, which plays the role of a Hubble-like length scale, encodes the central energy density and generates a repulsive core that prevents the occurrence of curvature singularities. 

The Hayward metric is given by:

\begin{equation}\label{f}
ds^{2} = -f(r)\, dt^{2} + \frac{dr^{2}}{f(r)} + r^{2} d\theta^{2} + r^{2}\sin^{2}\theta\, d\phi^{2},
\end{equation}
with
\begin{equation}\label{f1}
f_{\text{Haw}}(r)= 1 - \frac{2 M r^{2}}{r^{3} + 2 M \epsilon^{2}},
\end{equation}

Here, $M$ denotes the mass parameter. In the limit $\epsilon\to 0$, the metric (\ref{f1}) reduces to the Schwarzschild solution. The Hayward solution admits a critical mass $\overline{M}=(3\sqrt{3}/4)\,\epsilon$ and a corresponding radius $\overline{r}=\sqrt{3}\,\epsilon$ that determine its causal structure: for $M<\overline{M}$ no horizon is present; for $M=\overline{M}$ a single horizon appears at $r=\overline{r}$; and for $M>\overline{M}$ the Hayward BH exhibits two horizons.

The PFDM model was first proposed by Kiselev \cite{Kiselev:2003ah}, whose solution introduces a dark matter distribution modeled as a particular anisotropic fluid. Building on this idea, several works have subsequently explored different aspects and generalizations \cite{Li:2012zx}.

Since the Hayward space-time (\ref{f1}) can be described within the framework of nonlinear electrodynamics (NLED) coupled to Einstein gravity, one may interpret it as being sourced by a magnetic monopole whose magnetic charge is related to the Hayward parameters through $q_m^{3}=2\epsilon^{2}M$.

To obtain the Hayward black hole solution immersed in perfect fluid dark matter (PFDM), we consider the following action \cite{Tan:2025usr}:

\begin{equation}\label{action}
S=\int d^4x\sqrt{-g}\left[\frac{R}{16\pi}-\frac{\mathcal L(F)}{4\pi}+\mathcal L_{\mathrm{PFDM}}\right],
\end{equation}

where $R$ is the Ricci scalar, $\mathcal{L}(F)$ denotes the nonlinear electrodynamics Lagrangian (an arbitrary function of the electromagnetic invariant $F=(1/4)F_{\mu\nu}F^{\mu\nu}$), and $\mathcal{L}_{\mathrm{PFDM}}$ represents the Lagrangian density associated with perfect fluid dark matter.

By varying the action (\ref{action}), the Einstein field equations take the form:

\begin{equation}\label{EcEinst}
R_{\mu\nu}-\frac{1}{2}g_{\mu\nu}R = \frac{1}{4\pi}\left(\mathcal L_F F_{\mu\lambda}F_\nu{}^\lambda-g_{\mu\nu}\mathcal L\right) \pm
8\pi T_{\mu\nu}^{\mathrm{PFDM}}
\end{equation}

where $\mathcal L_F=\frac{d\mathcal L}{dF}$ and $T_{\mu\nu}^{(\text{PFDM})}$ represents the PFDM contribution. In the PFDM case, the energy--momentum tensor can be written as $T_{\mu}^{\nu}=\mathrm{diag}(-\rho, p_{r}, p_{\theta}, p_{\phi})$, where $\rho$ represents the dark matter energy density and $(p_{r}, p_{\theta}, p_{\phi})$ are the pressures \citep{Wang:2026sqr}.

The nonlinear electrodynamics Lagrangian for the Hayward space-time (\ref{f1}) is given by

\begin{equation}
\mathcal L(F)=\frac{3M}{q^{3}_{m}}\frac{(2q^{2}_{m}F)^{\frac{3}{2}}}{\left(1+(2q^{2}_{m}F)^{\frac{3}{4}}\right)^{2}}
\end{equation}

By considering the action (\ref{action}) and solving the Einstein field equations (\ref{EcEinst}), the metric function describing a Hayward black hole immersed in perfect fluid dark matter (Hayward--PFDM) \cite{Rizwan:2023ivp} is given by:

\begin{equation}\label{f2}
f(r)=1 - \frac{2 M r^{2}}{r^{3} + 2 M \epsilon^{2}}+\frac{\alpha}{r}\ln \left(\frac{r}{|\alpha|}\right).
\end{equation}

Here $\alpha$ is an integration constant that characterizes the strength of the dark matter distribution. Several studies have shown that the presence of PFDM can modify black hole space-times through the logarithmic term in the metric function \cite{Shaymatov:2020bso,Zhang:2020mxi,Abbas:2023pug,Rayimbaev:2021kjs}.

Depending on how the tensor $T_{\mu\nu}^{(\text{PFDM})}$ is incorporated into the Einstein equations (\ref{EcEinst}), the constant $\alpha$ may be taken to be positive or negative \cite{Das:2021otl} in order to satisfy the weak energy condition, leading to different BH solutions. This follows from the fact that the PFDM energy density is related to $\alpha$ as $\rho=\mp\alpha/(8\pi r^{3})$.

It is worth noting that, in the absence of PFDM ($\alpha\to 0$), the space-time described by Eq.~(\ref{f2}) reduces to the Hayward black hole (\ref{f1}). Conversely, when $\epsilon=0$, the solution recovers the Schwarzschild black hole immersed in PFDM \cite{Shaymatov:2020bso}.

The Hayward black hole (\ref{f1}) is regular everywhere. However, the inclusion of PFDM modifies this property, since the Hayward--PFDM solution (\ref{f2}) develops a singularity at $r=0$ due to the dark matter contribution, as evidenced by the divergence of the curvature scalar $R$ (\ref{R}), the Ricci invariant $R_{\mu\nu}R^{\mu\nu}$ (\ref{Ric}), and the Kretschmann scalar (\ref{K}). A similar behavior has been reported for a Bardeen black hole surrounded by dark matter \cite{Zhang:2020mxi}.

\begin{equation}\label{R}
R=\frac{24 M^2 \epsilon ^2 \left(4 M \epsilon ^2-r^3\right)}{\left(2 M \epsilon ^2+r^3\right)^3}-\frac{\alpha}{r^3}
\end{equation}

\begin{equation}\label{Ric}
R_{\mu\nu} R^{\mu\nu}= \frac{288 M^4 \epsilon^4 (5 r^6 - 4 M r^3 \epsilon^2 + 8 M^2 \epsilon^4)}{(r^3 + 2 M\epsilon^2)^6} -\frac{48 M^2 \alpha \epsilon^2 (2 r^3 + M \epsilon^2)}{r^3 (r^3 + 2 M \epsilon^2)^3}+\frac{5 \alpha^2}{2 r^6}
\end{equation}

and

\begin{eqnarray}\label{K}
&&K=R_{\mu \nu \rho \sigma}R^{\mu \nu \rho \sigma}=\frac{48 M^2 (r^{12} - 8 M r^9 \epsilon^2 + 72 M^2 r^6 \epsilon^4 - 
   16 M^3 r^3 \epsilon^6 + 32 M^4 \epsilon^8)}{(r^3 + 
  2 M \epsilon^2)^6}
 \nonumber \\
&& -\frac{8 M \alpha \left(-5 r^6 + 46 M r^3 \epsilon^2 + 4 M^2 \epsilon^4 + 
   6 (r^6 - 4 M r^3 \epsilon^2)\ln \left(\frac{r}{|\alpha|}\right)\right)}{(r^4 + 
  2 M r \epsilon^2)^3} 
  \nonumber\\
  && + \frac{\alpha^2 \left(13 + 
   4 \ln \left(\frac{r}{|\alpha|}\right)\left(-5 + 3 \ln \left(\frac{r}{|\alpha|}\right)\right)\right)}{r^6}
\end{eqnarray}

\subsection{Event horizons}

The horizons of the Hayward--PFDM space-time are determined by the positive roots of $f(r)=0$, with $f(r)$ given in Eq.~(\ref{f2}). For this analysis we introduce the rescaled variables $r\to r/|\alpha|$, $M\to M/|\alpha|$, and $\epsilon\to \epsilon/|\alpha|$ (these rescalings are used exclusively in the horizon analysis). The horizon radii are then obtained from the roots of

\begin{equation}\label{M}
2 M r^3-2 M r \epsilon^{2} -2 \tau M \epsilon^{2} \ln (r)-r^4-r^3 \tau \ln (r)=0.
\end{equation}

Here $\tau=\pm 1$ for $\alpha>0$ and $\alpha<0$, respectively.

It is evident that the number of horizons depends on the values of the Hayward--PFDM parameters. By applying the method used in \cite{Toshmatov:2015npp,Lopez:2021ujg}, one can determine the parameter ranges of $M$ and $\epsilon$ for which horizons exist.

From Eq.~(\ref{M}), $M$ can be parametrized as a function of $r$ and $\epsilon$ as

\begin{equation}\label{M1}
M(r,\epsilon^{2})=\frac{r^3 (r+\tau  \ln (r))}{2 \left(r^3-\tau  \epsilon^{2}  \ln (r)-r \epsilon^{2} \right)},
\end{equation}

then $M(r,\epsilon^{2})$ (\ref{M1}) has extrema ($dM(r,\epsilon)/dr=0$) for
$\epsilon^{2}(r)$ given by

\begin{equation}\label{eps}
\epsilon^{2}(r)=\frac{(r+ \tau) r^3}{3 (r + \tau\ln (r))^2}.
\end{equation}

The function $\epsilon^{2}(r)$ (\ref{eps}) exhibits a critical (minimum) value at $r_{\text{crit}}=1.09759$ (for $\alpha>0$) and at $r_{\text{crit}}=0.774652$ (for $\alpha<0$). A detailed analysis yields $\epsilon^{2}(1.09759)=(\epsilon^{2}_{\text{crit}})_{\alpha>0}=0.65209$ and $\epsilon^{2}(0.774652)=(\epsilon^{2}_{\text{crit}})_{\alpha<0}=-0.0329142$. However, since $\epsilon^{2}$ must be positive, we take $(\epsilon^{2}_{\text{crit}})_{\alpha<0}=0$.

The corresponding critical masses $M_{\text{crit}}$ are obtained by evaluating Eq.~(\ref{M1}) at $r=r_{\text{crit}}$ and $\epsilon^{2}=\epsilon^{2}_{\text{crit}}$, giving $(M_{\text{crit}})_{\alpha>0}=1.44227$ (minimum) and $(M_{\text{crit}})_{\alpha<0}=0.514997$ (minimum). In summary, for $\epsilon^{2}_{\text{crit}}\leq \epsilon^{2}$ and $M_{\text{crit}}\leq M$, the Hayward--PFDM space-time has horizons.

We next consider the extremal case, where $df(r)/dr=0$ and $f(r)=0$ are satisfied simultaneously; in this situation one obtains the following relation:

\begin{equation}\label{ME}
M(r)= \frac{3 (\tau  \ln (r)+r)^2}{6 \tau  \ln (r)+4 r-2 \tau }.   
\end{equation}

\begin{figure}[ht]
\begin{center}
\includegraphics [width =0.49 \textwidth ]{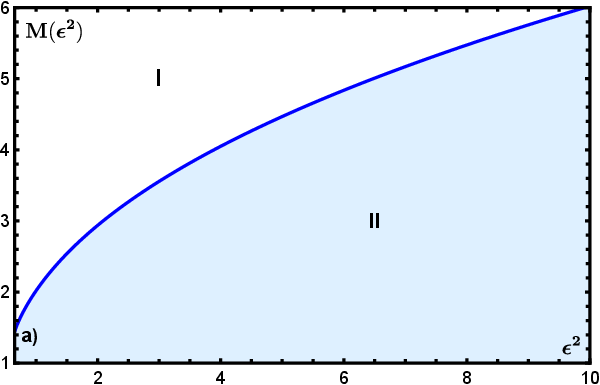}
\includegraphics [width =0.49 \textwidth ]{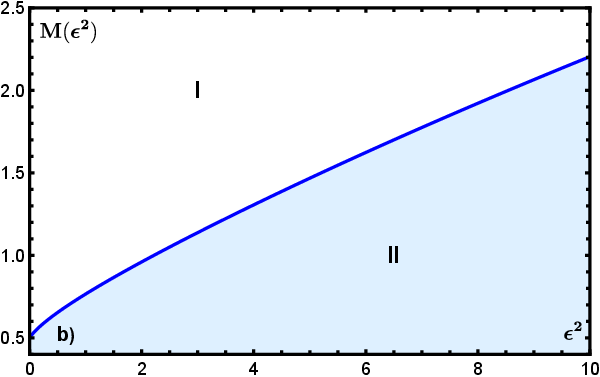}
\end{center}
\caption{The figures show the behavior of $M$ as a function of $\epsilon^{2}$. (a) For $\alpha>0$, region I contains the values of $M$ and $\epsilon^{2}$ for which the Hayward--PFDM black hole has three horizons; in region II it has one horizon. The boundary between regions I and II represents the extremal Hayward--PFDM configuration (two horizons). (b) For $\alpha<0$, region I contains the values of $M$ and $\epsilon^{2}$ for which the Hayward--PFDM black hole has two horizons; in region II it has no horizons. The boundary between regions I and II represents the extremal Hayward--PFDM configuration (one horizon). }\label{Fig1}
\end{figure}

In Fig.~\ref{Fig1}, the dependence of $M$ on $\epsilon^{2}$ is obtained from Eqs.~(\ref{eps}) and (\ref{ME}). For a fixed value of $\tau$, each value of $r$ uniquely determines both the mass parameter $M$ via Eq.~(\ref{ME}) and the parameter $\epsilon^{2}$ via Eq.~(\ref{eps}). Consequently, these equations define a one-to-one mapping between $M$ and $\epsilon^{2}$, from which the function $M(\epsilon^{2})$ is obtained.

For $\alpha>0$, the number of horizons increases and the Hayward--PFDM black hole may exhibit up to three horizons (region I in Fig.~\ref{Fig1}a)), a feature also reported for black holes surrounded by dark energy \cite{Ramirez:2021ibk},\cite{Pedraza:2021hzw},\cite{Fernando:2012ue}. The boundary between regions I and II corresponds to the extremal configuration with two horizons, while region II corresponds to solutions with a single horizon.

Figure~\ref{Fig1}b) shows the case $\alpha<0$. In region I the Hayward--PFDM space-time has two horizons, whereas in region II it has no horizons. The boundary between regions I and II corresponds to the extremal Hayward--PFDM configuration (one horizon). In this case, depending on the values of $M$ and $\epsilon^{2}$, the number of horizons can decrease from two to one; thus, introducing PFDM does not increase the number of horizons for $\alpha<0$.

A comparison of the critical masses shows that a positive PFDM parameter ($\alpha>0$) increases the critical mass. Specifically, the ordering satisfies
$\overline{M}<\left(M_{\mathrm{crit}}\right)_{\alpha<0}<\left(M_{\mathrm{crit}}\right)_{\alpha>0}$.

\subsection{Effective metric}\label{SubEff}

It is important to emphasize that, although the PFDM contribution modifies the background space-time geometry, the nonlinear electrodynamics sector remains present and continues to influence photon propagation. In particular, unlike in linear electrodynamics, where photons follow null geodesics of the background metric $g_{\mu\nu}$, the nonlinear nature of the electromagnetic field causes photons to propagate along null geodesics of an effective geometry characterized by the metric tensor $\hat{g}_{\mu\nu}$, which depends on the underlying NLED theory. Therefore, the inclusion of PFDM does not eliminate the effects of NLED on photon trajectories, making it necessary to employ the corresponding effective metric. This effective geometry can be derived by analyzing the propagation of electromagnetic field discontinuities across characteristic hypersurfaces \cite{Novello:1999pg}. Accordingly, the gradient of the characteristic surface, $S{,\mu}$, satisfies the following equation:

\begin{equation}\label{geff}
\left(\mathcal{L}_{F}g^{\mu\nu} - \mathcal{L}_{FF}F^{\mu}{}_{\alpha}F^{\alpha\nu}\right)S_{,\mu}S_{,\nu}=\hat{g}_{\mu\nu}S_{,\mu}S_{,\nu}=0.
\end{equation}

The derivation of effective metrics in NLED has been extensively discussed in \cite{Breton:2016mqh}. In particular, for the Hayward space-time, the corresponding effective metric was obtained in \cite{Becerril:2020fek} , yielding the following expressions:

\begin{equation}
g_{tt}=\hat{g}_{tt}  \;\;\;\;\;\ g_{rr}=\hat{g}_{rr}  \;\;\;\;\;\  \hat{g}_{\theta \theta}=\frac{1}{G_{m}(r)}g_{\theta \theta}  \;\;\;\;\;\  \hat{g}_{\phi\phi}=\frac{1}{G_{m}(r)}g_{\phi\phi}
\end{equation}

where

\begin{equation}
G_{m}(r)=\frac{2 r^3 - 5 M \epsilon^2}{r^3 + 2 M \epsilon^2} 
\end{equation}

This distinction will be taken into account in the analysis of the photosphere, frequency shifts, and the black hole shadow.

\section{Effective potential for test particles}

As a first step in determining the frequency shifts of particles orbiting a black hole, we analyse the effective potential in order to identify the conditions required for the existence of stable orbits.

For this purpose, we consider a static and axially symmetric space-time, described by Eq.~(\ref{f}) in spherical coordinates, and study the motion of test particles through the geodesic equations obtained from the Lagrangian formalism:

\begin{equation}
\mathcal{L} = g_{\mu\nu} U^{\mu} U^{\nu},
\end{equation}

where  $U^{\mu} = \frac{dx^{\mu}}{d \eta}$ is the four-velocity and $\eta$ is an affine parameter. The condition $\mathcal{L}=0$ corresponds to null geodesics (photons), whereas $\mathcal{L}=-1$ characterizes the trajectories of massive particles. 

The conserved quantities associated with the motion of massive particles are the energy $E$ and the angular momentum $L$, which are given by:

\begin{equation}
E = f(r)U^{t}, \qquad L = r^{2} \sin \theta U^{\phi}.
\end{equation}

For equatorial orbits ($\theta = \pi/2$), the radial equation of motion for a massive particle in the static space-time can be written as $(U^{r})^{2}+V_{\text{eff}}=E^{2}$, where $V_{\text{eff}}$ denotes the effective potential governing the radial motion:

\begin{equation}\label{Veff}
V_{\text{eff}}= \left(1+ \frac{L^{2}}{r^{2}}\right) f(r).
\end{equation}

For the Hayward--PFDM case, the effective potential describing the radial motion of massive particles is given by:

\begin{equation}
V_{\text{eff}}= \frac{\left(L^2+r^2\right) \left(r^4+A+B\right)}{r^3 \left(2 M \epsilon ^2+r^3\right)},
\end{equation}

where $A= \alpha  \left(2 M \epsilon ^2+r^3\right) \ln \left(\frac{r}{| \alpha | }\right)$ and $B= 2M r(\epsilon^2 -r^2)$.

\begin{figure}[ht]
\begin{center}
\includegraphics [width =0.495 \textwidth ]{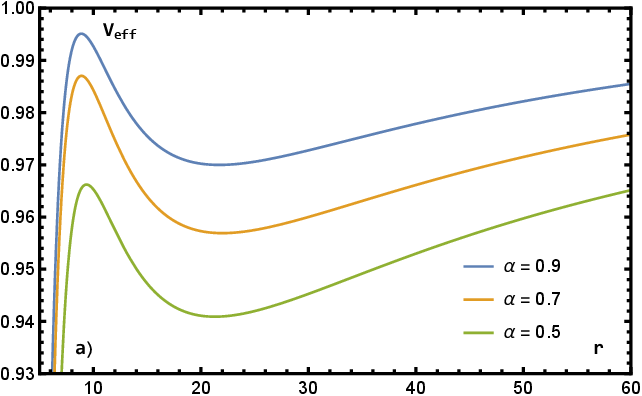}
\includegraphics [width =0.495 \textwidth ]{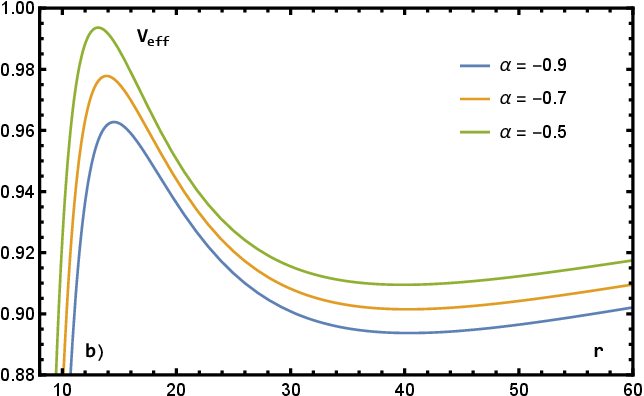}
\end{center}
\caption{a) Behavior of the effective potential $V_{\text{eff}}$ for different values of the PFDM parameter $\alpha>0$ with angular momentum $L=6.8$.  b) Behavior of the effective potential $V_{\text{eff}}$ for different values of the PFDM parameter $\alpha<0$ with angular momentum $L=14$. In both panels, we consider $M=2.7$ and $\epsilon=0.85$. }\label{Fig1.1}
\end{figure}

In Fig.~\ref{Fig1.1}a), we show the behavior of the effective potential $V_{\text{eff}}$, Eq.~(\ref{Veff}), for the Hayward--PFDM space-time for different values of $\alpha>0$. The effective potential exhibits both a maximum and a minimum, indicating the presence of unstable and stable circular orbits, respectively. The plot also shows that the motion of test particles is noticeably affected by the PFDM parameter. In particular, as $\alpha$ increases, the maximum of the effective potential shifts toward the horizon, whereas the minimum is displaced in the opposite direction.

In Fig.~\ref{Fig1.1}b), we show $V_{\text{eff}}$ for different values of $\alpha<0$. The behavior is similar to that displayed in Fig.~\ref{Fig1.1}a), exhibiting both a maximum and a minimum. However, the maximum values of the potential are lower for $\alpha<0$, indicating that the PFDM parameter significantly modifies the effective potential governing particle motion.

We note that the existence of extrema (both maxima and minima) of the effective potential is constrained by the angular momentum $L$. In particular, the formation of minima for $\alpha > 0$ requires smaller values of $L$ than those corresponding to $\alpha < 0$. 

\section{Orbits around a Hayward-PFDM}

Before analysing the red/blue shifts of photons emitted by massive particles orbiting the Hayward--PFDM black hole, it is necessary to examine the possible orbits. We restrict our analysis to circular orbits. Such trajectories must satisfy $V_{\text{eff}}(r)=E^{2}$ and $V'_{\text{eff}}(r)=0$, where the prime denotes differentiation with respect to the radial coordinate $r$. For a circular orbit of radius $r_{c}$ to be stable, it is additionally required that $V''_{\text{eff}}(r_{c})>0$, ensuring that small radial perturbations do not drive the particle away from the circular trajectory.

Therefore, we can derive the following expressions \cite{Becerril:2016qxf} for the energy and angular momentum of a particle in circular orbits:

\begin{equation} \label{EL}
    E^{2}  = \frac{2f^2(r)}{2f(r) - rf^{'}(r)}, \qquad L^{2}  = \frac{r^{3}f^{'}(r)}{2f(r) - rf^{'}(r)}.
\end{equation}

For the Hayward--PFDM space-time, Eq.~(\ref{EL}) becomes:

\begin{equation}\label{E}
E^2 = \frac{2 \left(A+B+r^4\right)^2}{r \left(3 A \left(2 M \epsilon ^2+r^3\right)-r^6 (\alpha +6 M-2 r)+4 M \epsilon ^2 (2 r-\alpha ) \left(M \epsilon ^2+r^3\right)\right)},
\end{equation}

and

\begin{equation}\label{L}
L^2 = \frac{r^2 \left(-A \left(2 M \epsilon ^2+r^3\right)+M^2 \left(4 \alpha  \epsilon ^4-8 r^3 \epsilon ^2\right)+2 M \left(r^6+2 \alpha  r^3 \epsilon ^2\right)+\alpha  r^6\right)}{3 A \left(2 M \epsilon ^2+r^3\right)-r^6 (\alpha +6 M-2 r)+4 M \epsilon ^2 (2 r-\alpha ) \left(M \epsilon ^2+r^3\right)}.
\end{equation}

In particular, for $\epsilon = 0$, the expressions for the energy (\ref{E}) and angular momentum (\ref{L}) reduce to those of the Schwarzschild black hole immersed in PFDM (see (\ref{EL})). 

\begin{equation}\label{EL}
E^2 =\frac{2 \left(\alpha  \ln\left(\frac{r}{| \alpha | }\right)-2 M+r\right)^2}{r \left(-\alpha +3 \alpha  \ln \left(\frac{r}{| \alpha | }\right)-6 M+2 r\right)}, \qquad L^{2}=\frac{r^2 \left(\alpha -\alpha  \ln \left(\frac{r}{| \alpha | }\right)+2 M\right)}{-\alpha +3 \alpha  \ln \left(\frac{r}{| \alpha | }\right)-6 M+2 r}
\end{equation}

Furthermore, in the limit $\alpha \to 0$, the expressions in (\ref{EL}) reduce to the corresponding Schwarzschild expressions.

Circular orbits of massive particles do not exist for arbitrary values of $r$. In particular, when the energy given by Eq.~(\ref{E}) diverges, the corresponding trajectory approaches a photon orbit. However, in nonlinear electrodynamics, photon trajectories are governed by an effective optical metric $\hat{g}_{\mu\nu}$ rather than by the background geometry (see Subsection~\ref{SubEff}). Therefore, the photon-sphere radius $r_{\text{ph}}$ must be determined from null circular geodesics of the effective metric and is given by:

\begin{equation}\label{Ph}
(2f(r) - rf^{'}(r))G_{m}(r)=f(r)rG^{'}(r)_{m}
\end{equation}

The marginally bound circular orbits ($r_{\text{mb}}$) are characterized by the condition $E = 1$, while the innermost stable circular orbits (ISCO) are determined by the requirement $V''_{\text{eff}}(r)=0$ in $r=r_{\text{ISCO}}$ given by:

\begin{equation}\label{V2}
 rf(r)f^{''}(r)+3f(r)f^{'}(r)-2r(f^{'}(r))^{2} =0,   
\end{equation}

For the Hayward--PFDM metric, Eq.~(\ref{V2}) leads to the following explicit condition, Eq.~(\ref{ISCO}):

\begin{eqnarray}\label{ISCO}
&8Mr^{3}\left(r^{3}-4M\epsilon^{2}\right)\alpha P_{0}
+2\alpha^{3}P_{0}^{3}
+2Mr^{4}P_{2}+3AP_{0}^{2}\ln\left(\frac{r}{|\alpha|}\right) \nonumber\\
&+P_{1}(r-4\alpha)\alpha\ln\!\left(\frac{r}{|\alpha|}\right) -12Mr^{3}\alpha\left(r^{3}-2M\epsilon^{2}\right)
\left(r^{3}+5M\epsilon^{2}\right)
\ln \left(\frac{r}{|\alpha|}\right)=0.
\end{eqnarray}

where $P_{1}=8 M^3 r^3 \epsilon ^4+12 M^2 r^3 \epsilon ^4+6 M r^6 \epsilon ^6+r^9$, $P_{2}=6Mr^{5}-r^{6}-22Mr^{3}\epsilon^{2}+32M^{2}\epsilon^{4}$, and $P_{0}=2 M\epsilon^2+r^3$.

\begin{figure}[ht]
\begin{center}
\includegraphics [width =0.49 \textwidth ]{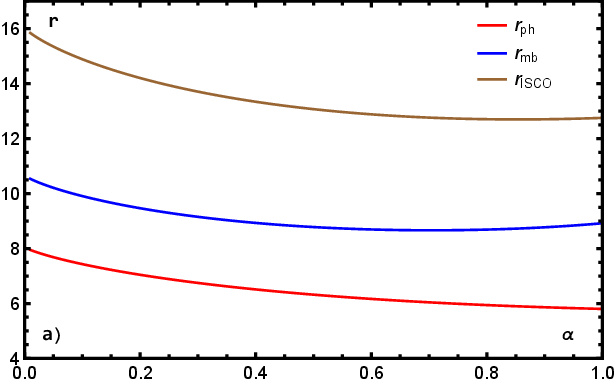}
\includegraphics [width =0.49 \textwidth ]{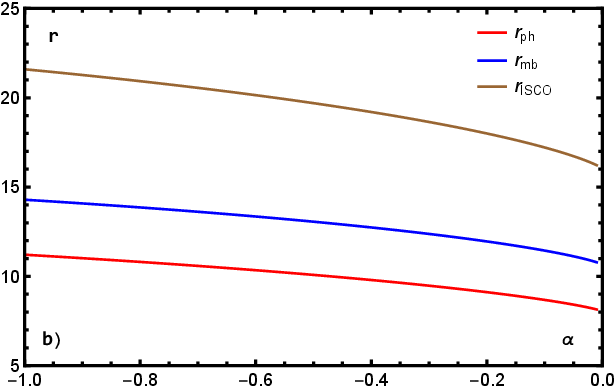}
\end{center}
\caption{Behavior of the photon sphere radius $r_{\mathrm{ph}}$, the marginally bound orbit radius $r_{\mathrm{mb}}$, and the innermost stable circular orbit radius $r_{\mathrm{ISCO}}$ as functions of the PFDM parameter $\alpha$, for fixed values of $M=2.7$ and $\epsilon=0.85$. 
Panel a) corresponds to $\alpha>0$, while panel b) corresponds to $\alpha<0$.}\label{Fig2}
\end{figure}

A comparison among the characteristic radii is shown in Fig.~\ref{Fig2}: the photon sphere $r_{\text{ph}}$, the marginally bound orbit $r_{\text{mb}}$, and the ISCO radius $r_{\text{ISCO}}$. These radii depend on the PFDM parameter $\alpha$.

In Fig.~\ref{Fig2}a) ($\alpha>0$), all radii decrease as $\alpha$ increases, preserving the ordering $r_{\text{ph}}<r_{\text{mb}}<r_{\text{ISCO}}$. This indicates that PFDM shifts the region of stable motion inward, which can be interpreted as an enhancement of the effective gravitational attraction at intermediate scales and thus requires smaller orbital radii to sustain stable circular motion.

In contrast, Fig.~\ref{Fig2}b) ($\alpha<0$) exhibits the same ordering $r_{\text{ph}}<r_{\text{mb}}<r_{\text{ISCO}}$, but all radii increase with increasing $|\alpha|$. In this regime, the characteristic radii decrease as $\alpha\to 0$.

In the limit $\alpha\to 0$, both branches smoothly converge to the pure Hayward solution, confirming that PFDM acts as a continuous deformation of the space-time. Therefore, the parameter $\alpha$ governs the displacement of circular orbits and plays a central role in determining the orbital dynamics and the associated observable effects.

\section{The red/blue shifts for photon}

The relationship between the red/blue shifts of photons emitted by massive particles (such as stars or gas) moving along stable geodesics around a black hole was first highlighted in \cite{Herrera-Aguilar:2015kea}. Subsequently, the same method was developed for static space-times in \cite{Becerril:2016qxf}. 

To proceed with the analysis, we consider a photon emitted by particles orbiting the Hayward--PFDM black hole on stable circular orbits of radius $r_{c}$ in the equatorial plane ($\theta=\pi/2$). The photon four-momentum is denoted by $K^{\mu}$ and satisfies the null condition $K^{\mu}K_{\mu}=0$. Moreover, the photon energy and angular momentum are conserved quantities, given by:

\begin{equation}\label{ELph}
 E_{ph} = f(r)\,K^{t}, 
\qquad 
L_{ph} = \frac{r^{2}}{G_{m}}\,K^{\phi}.
\end{equation}

The frequency shift $z$ associated with the photon, measured by an emitter ($e$) and a detector ($d$), is defined as $1+z=\omega_{e}/\omega_{d}$, where the measured frequencies are

\begin{equation}
\omega_{e}=-K_{\mu}U^{\mu}_{e}, \qquad \omega_{d}=-K_{\mu}U^{\mu}_{d}.
\end{equation}

For circular geodesic motion in the equatorial plane, $U_{r}=0$ and $U_{\theta}=0$, so that

\begin{equation}
\omega_{e}=E_{ph}U^{t}_{e}-L_{ph}U^{\phi}_{e}, \qquad \omega_{d}=E_{ph}U^{t}_{d}-L_{ph}U^{\phi}_{d}.
\end{equation}

Therefore, the frequency shift can be written as

\begin{equation}
1+z=\frac{U_{e}^{t}-b_{e}U^{\phi}_{e} }{U_{d}^{t}-b_{d}U^{\phi}_{d}},
\end{equation}
 
From $K^{\mu}K_{\mu}=0$ with $K^{\phi}=K^{\theta}=0$, one obtains the impact parameter $b=\sqrt{-\hat{g}_{\phi\phi}/g_{tt}}$. Owing to the conservation of energy and angular momentum, the impact parameter remains invariant along the trajectory, implying $b_{e}=b_{d}$.

Observational frequency shifts are commonly expressed in terms of the kinematic shift,
$z_{\text{kin}}=z-z_{c}$, where $z_{c}$ corresponds to a photon emitted by a static particle with $b=0$.

If the detector is located far away from the black hole ($r\to\infty$), its four-velocity becomes $U^{\mu}_{d}=(1,0,0,0)$.Under these conditions, the frequency shift can be expressed as Eq.~(\ref{z}), incorporating the effects of NLED on photon propagation \cite{Lopez:2020ovf, Becerril:2020fek}:

\begin{equation}\label{z}
z_{\text{kin}} = U^{\phi}_{e}b_{e \pm} = \pm  \sqrt{\frac{1}{G_{m}(r)}} \sqrt{ \frac{rf'(r)}{f(r) \left( 2 f(r) - r f'(r) \right)}}.
\end{equation}

The positive sign corresponds to the redshift ($z_{\text{red}}$), while the negative sign corresponds to the blueshift ($z_{\text{blue}}$). In terms of the Hayward--PFDM metric, Eq.~(\ref{z}) becomes:

\begin{equation}
   z_{\text{kin}}= \pm \sqrt{\frac{r^3 + 2 M \epsilon^2}{2 r^3 - 5 M \epsilon^2}} \sqrt{\frac{L^{2} \left(r^3+2 M \epsilon^2\right)}{r(r^4+A+B)}}.
\end{equation}

\begin{figure}[ht]
\begin{center}
\includegraphics [width =0.49 \textwidth ]{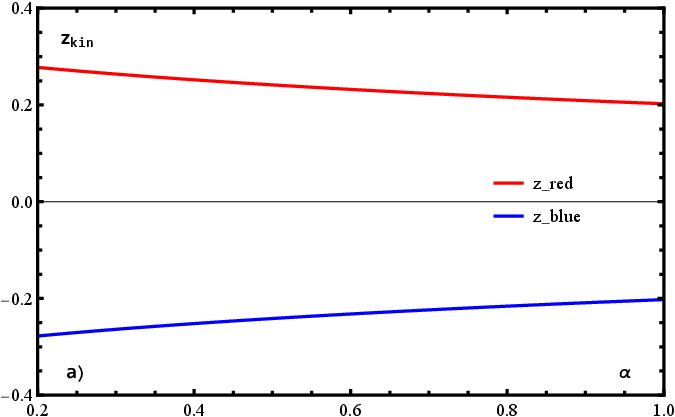}
\includegraphics [width =0.49 \textwidth ]{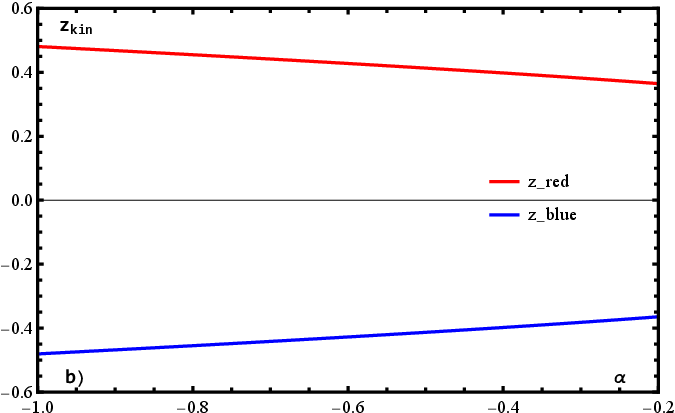}
\end{center}
\caption{Behavior of $z_{\text{kin}}$ as a function of $\alpha$. Panel (a) corresponds to $\alpha>0$, while panel (b) corresponds to $\alpha<0$. In both panels we use $M=2.7$, $\epsilon=0.85$, and $r=25$.}\label{Fig3}
\end{figure}

Choosing appropriate emitter radii $r_e \geq r_{ISCO}$ within the physical domain of stable time-like circular orbits, and considering the parameter ranges displayed in Fig.~\ref{Fig1}, we analyze the kinematic frequency shift $z$ for the Hayward-PFDM black hole. The ISCO radius provides a natural choice for the inner edge of the emitting disk, while larger fixed radii allow for controlled comparisons. The results for positive and negative values of the PFDM parameter $\alpha$ are presented in Fig.~\ref{Fig3}.

For $\alpha>0$ (Fig.~\ref{Fig3} a)), the redshift decreases as $\alpha$ increases, whereas the blueshift becomes less negative. This means that, in this branch, the PFDM contribution reduces the magnitude of the frequency shift as $\alpha$ grows. 

For $\alpha<0$ (see Fig.~\ref{Fig3} b),  as $\alpha$ approaches zero, the redshift decreases and the blueshift increases. In particular, negative values of $\alpha$ enhance the absolute value of the blueshift, indicating a stronger kinematic contribution to the variation in photon frequency. However, the redshift remains above the value obtained for the pure Hayward black hole, whose minimum is reached at $\alpha=0$.

Thus, Fig. \ref{Fig3} shows that the sign of $\alpha$ has a direct impact on the observable frequency shifts. Positive values of $\alpha$ tend to reduce the magnitude of both red/blue shifts as $\alpha$ increases, while negative values produce larger frequency shift magnitudes. This behavior indicates that red/blue shifts measurements from particles orbiting compact objects may provide a possible observational signature of the PFDM contribution in the space-time.

\section{Shadow of the Black Hole}

The results of the Event Horizon Telescope (EHT) collaboration have significantly advanced the study of black hole shadows. In particular, the images of M87* (2019) \cite{EventHorizonTelescope:2019dse} and Sgr  A* \cite{EventHorizonTelescope:2022wkp} \citep{Vagnozzi:2022moj} provide direct observational evidence of the space-time geometry near these objects. The black hole shadow corresponds to the dark region formed by photon capture, whose boundary is determined by unstable spherical photon orbits. For a distant observer in a static and spherically symmetric space-time, the shadow radius can be obtained within the optical approximation, where it coincides with the critical impact parameter $b_c$, as;

\begin{equation}
r_{sh}=b_{c} = \sqrt{\frac{r_{ph}^{2}}{G_{m}(r_{ph})f(r_{ph})}},
\end{equation}

To confront the Hayward–PFDM black hole with observational data from Sgr A*, one employs the mass-to-distance ratio, together with a calibration factor that relates the observed emission ring to the actual shadow size. This factor is expected to be close to unity under reasonable astrophysical assumptions. Consequently, one can define the relative deviation $\delta$ with respect to the Schwarzschild shadow radius:

\begin{equation}
(r_{sh})_{Schw}=3\sqrt{3}M(1 + \delta),
\end{equation}

where $M$ denotes the Schwarzschild mass. Assuming a Gaussian distribution, the corresponding constraints on $\delta$ are $-0.125 \lesssim \delta \lesssim 0.005$ at the $1\sigma$ confidence level and $-0.19 \lesssim \delta \lesssim 0.07$ at the $2\sigma$ confidence level.

\begin{figure}[ht]
\begin{center}
\includegraphics [width =0.49 \textwidth ]{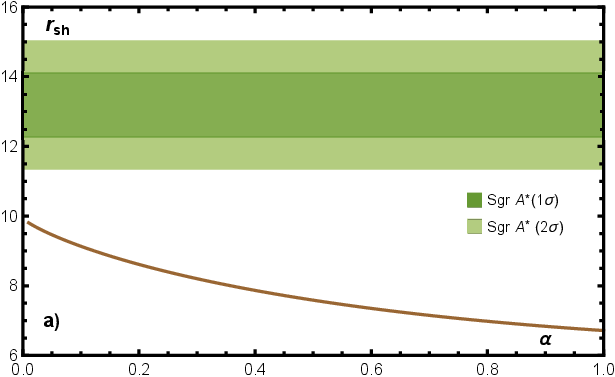}
\includegraphics [width =0.49 \textwidth ]{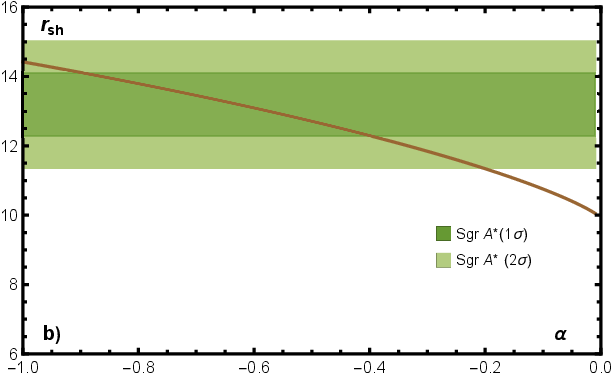}
\end{center}
\caption{Shadow radius of the Hayward-PFDM black hole as a function of the PFDM parameter $\alpha$: (a) $\alpha>0$ and (b) $\alpha<0$. The results are compared with the shadow radius of the Schwarzschild BH. The dark and light green bands represent the observational constraints at the $1\sigma$ and $2\sigma$ confidence levels, respectively, derived from the Sgr A$^{*}$ black hole observations. The parameters are fixed to $M=2.7$ and $\epsilon=0.85$.}\label{Fig4}
\end{figure}

The behavior of the shadow radius shown in Fig.~\ref{Fig4} reveals its dependence on the PFDM parameter $\alpha$. For $\alpha>0$ (Fig.~\ref{Fig4} a)), the shadow radius decreases monotonically as $\alpha$ increases, remaining below the observational bounds inferred from the EHT observations of Sgr A$^{*}$. This behavior indicates that positive PFDM contributions reduce the effective size of the photon capture region, shifting the photon sphere toward smaller radii.

In contrast, for $\alpha<0$ (Fig.~\ref{Fig4} b)), the shadow radius increases as $|\alpha|$ increases, reaching values larger than the Schwarzschild shadow radius for sufficiently negative $\alpha$. As $\alpha \rightarrow 0$, the shadow radius decreases and gradually approaches the pure Hayward limit. Moreover, the shadow radius falls within the observational bounds for certain negative values of $\alpha$.

Therefore, for the parameters considered, only a restricted range of negative values of the PFDM parameter is compatible with the observational constraints of Sgr A$^{*}$ at the $1\sigma$ and $2\sigma$ confidence levels.

\section{Conclusions}

In this paper, we have investigated the red/blue shifts of photons emitted by massive particles moving along stable circular orbits around Hayward black hole immersed in perfect fluid dark matter.
The horizon analysis revealed that the PFDM parameter changes the number of horizons. For $\alpha>0$, the Hayward-PFDM black hole may exhibit up to three horizons, whereas for $\alpha<0$ admits at most two horizons. In both cases, the critical values of the parameters depend strongly on the PFDM contribution, showing that the surrounding dark matter distribution modifies the black-hole geometry. 

The effective potential analysis showed the existence of stable and unstable circular orbits for both positive and negative values of $\alpha$. In particular, the PFDM parameter changes the location of the extrema of the effective potential, indicating that the motion of test particles is sensitive to the dark matter environment.

We also found that the characteristic radii associated with circular motion, namely the photon sphere $r_{ph}$, the marginally bound orbit $r_{mb}$, and the ISCO radius $r_{ISCO}$, preserve the ordering $r_{\rm ph}<r_{\rm mb}<r_{\rm ISCO}$,
although their magnitudes depend on the sign and value of $\alpha$.

For nonzero values of $\alpha$, the PFDM contribution modifies both the red/blue shifts, with different behaviors depending on the sign of the parameter. Positive values of $\alpha$ tend to reduce the magnitude of the shifts as $\alpha$ increases. These results indicate that the PFDM environment produces measurable modifications in the propagation of photons emitted near black holes.

In particular, the dependence of the frequency shifts on the sign and magnitude of $\alpha$ suggests that these observables may help to distinguish different dark matter configurations surrounding compact objects. 

Finally, the shadow of the Hayward–PFDM black hole was analyzed, revealing that the PFDM parameter significantly affects its size. Positive values of $\alpha$ decrease the shadow radius, whereas negative values generally lead to larger shadow sizes. Comparison with the observations of Sagittarius A* by the Event Horizon Telescope Collaboration indicates that only a restricted range of negative values of the PFDM parameter is consistent with the observational bounds, suggesting that black hole shadow measurements may provide constraints on the PFDM contribution.

\section*{ACKNOWLEDGMENT}

L. A. L\'opez acknowledges the support of SECIHTI-SNII, Mexico. Laura Marlene\'opez gratefully acknowledges the SECIHTI Master's Fellowship (CVU No. 2088339).

\bibliographystyle{unsrt}

\bibliography{bibliografia}

\end{document}